\documentclass [12pt]{iopart}
\def\veps{\varepsilon}

\begin{document}
\jl{2}
\letter{Breit interaction in heavy atoms}
\author{M G Kozlov\dag\ddag,
        S G Porsev\dag, and
        I I Tupitsyn\S}
\address{\dag\, Petersburg Nuclear Physics Institute,
        188350, Gatchina, Russia}
\address{\ddag\, E-mail: \tt mgk@MF1309.spb.edu}
\address{\S\, St.~Petersburg State University, Petrodvorets, Russia}
\date{\today}
\submitted
\begin{abstract}
In this letter we discuss how to include Breit interaction in
calculations of low-energy properties of heavy atoms in accurate
and effective way. In order to illustrate our approach we give some
results for Cs and Tl.

\end{abstract}


\paragraph{Introduction.}
It is well known that in high accuracy atomic calculations one may
need to go beyond the Coulomb approximation for the two-electron
interaction. The usual next step is to include the Breit
interaction. A number of commonly used atomic packages, such as
GRASP package~\cite{P}, allows to do this routinely. However, in
many precision calculations one needs to account for different
corrections, which are not included in standard packages. Thus, it
may be important to analyze the role of the Breit interaction on
different stages of atomic calculations. Below we present several
simple observations which may help to include Breit interaction in
atomic calculations in an optimal way.

Breit interaction between electrons 1 and 2 has the form
\begin{eqnarray}
\fl     V_{\rm B} \equiv V_{\rm B}^1 + V_{\rm B}^2
        = - \frac{\vec{\alpha}_1 \cdot \vec{\alpha}_2} {r_{12}}
        + \frac{1}{2} \left\{
        \frac{\vec{\alpha}_1 \cdot \vec{\alpha}_2} {r_{12}}
        -\frac{(\vec{r}_{12} \cdot \vec{\alpha}_1)
        (\vec{r}_{12} \cdot \vec{\alpha}_2)}{r_{12}^3}
        \right\},
\label{II_3}
\end{eqnarray}
where $\vec{\alpha}_i$ are Dirac matrices and $r_{12}$ is the
distance between the electrons. Equation \eref{II_3} can be readily
derived from the general expression for the interaction via
transverse photon in the Coulomb gauge by expanding it in powers of
the photon frequency (see, for example, \cite{LNL}). The operator
$V_{\rm B}^1$ is called the magnetic (or Gaunt) term and the
operator $V_{\rm B}^2$ is called the retardation term \cite{G}. The
magnetic term of the Breit operator is known to dominate in atomic
calculations \cite{G1} and it is much simpler than the whole Breit
operator. In present work we will neglect the retardation term and
we will check the accuracy of this approximation by comparison to
the resent calculation by Derevianko \cite{Der} where the whole
Breit operator was used. Note also, that some technical details of
atomic calculations with retardation term were recently discussed
in \cite{RH} (see also references therein).

\paragraph{Breit interactions in Dirac-Fock calculations.}
Here, we are interested in the calculations of the low-energy
properties of heavy atoms. In such calculations all electrons are
usually divided into two parts: the core electrons and the valence
ones. Correspondingly, the interaction between electrons is reduced
to the interaction between the valence electrons and the
interaction of the valence electrons with atomic core. The latter
is described by the Hartree-Fock potential which includes direct
and exchange terms. In this case the following observations can be
made.

\begin{enumerate}
\item
Breit correction to the interaction between the valence electrons
is of the order of $\alpha^2$ ($\alpha$ is the fine structure
constant) which is usually below the accuracy of modern
calculations of heavy atoms.

\item
Breit correction to the direct term of the Hartree-Fock potential
turns to zero if the core includes only closed shells. Indeed, the
vertex of the Breit interaction includes Dirac matrix
$\vec{\alpha}$, which is averaged to zero when summation over the
closed shell is done.

\item
Breit correction to the exchange interaction of the valence
electron with the core does not turn to zero and is of the order of
$R^2$, where $R$ is the overlap integral between the upper
component of a valence orbital and the lower component of a core
one. The largest integrals $R$ correspond to the innermost core
orbitals, where small components are of the order of $\alpha Z$.
Thus, the dominant Breit correction is the one to the exchange core
potential.

\item
The exchange interaction between valence electrons and the
innermost core electrons is significantly screened if the core
relaxation is allowed. Therefore, it is very important that Breit
correction to the core potential is calculated self-consistently.
In some cases the core relaxation can reduce the final Breit
correction to the valence energies by an order of magnitude (see
the results for Cs below).
\end{enumerate}

We are not going to give all technical details of the calculations
here, but there are few points to be mentioned, at least briefly.
Above we have argued that the only important contribution of the
Breit interaction comes from the exchange with the core. The
corresponding matrix element can be written in a form:
\begin{eqnarray}
\fl     \langle f |V_{\rm B}^{\rm core}| i \rangle
        = - \sum_{c \in \rm core}
        \langle f,c|V_{\rm B}|c,i \rangle
\nonumber \\
        = - \delta_{j_f,j_i} \delta_{m_f,m_i}
        \sum_k \sum_{c \in \rm core} (2j_c+1)
        \left(\begin{array}{ccc} j_f & j_c & k \\
                                 -\case{1}{2} & \case{1}{2} & 0
        \end{array} \right)^2
        R^k_{f,c,c,i},
\label{II_5}
\end{eqnarray}
where $R^k_{f,c,c,i}$ is the radial integral of the Breit
interaction for multipolarity $k$.

Interaction \eref{II_5} can be included in calculation either
perturbatively or self-consistently. In the former approach, the
first order correction to the energy of the valence electron $v$ is
simply
$ \delta \veps_v = \langle v |V_{\rm B}^{\rm core}| v \rangle$.
In the self-consistent approach, the potential $V_{\rm B}^{\rm
core}$ should be included in the Dirac-Fock (DF) equations. These
equations will then give new set of orbitals and energies
$\{\tilde{\varphi}_n,\tilde{\veps}_n\}$ and
$ \delta \veps_v = \tilde{\veps}_v-\veps_v.$

There are two things one has to keep in mind when solving the
Dirac-Fock-Coulomb-Breit (DFCB) equations. The first is that Breit
approximation is not completely relativistic and, thus, some
caution may be necessary. For example, one can use projectors to
the positive energy states. Technically that can be done with the
help of the kinetic balance condition for the small components of
the Dirac orbitals. Second, if we include Breit interaction in DF
equations, the resultant corrections are not linear in Breit
interaction. It is not difficult to eliminate the higher orders in
Breit interaction with the help of a scaling parameter $\lambda$:
one can substitute $V_{\rm B}$ with $\lambda V_{\rm B}$ and then
calculate $\lim_{\lambda \rightarrow
0}(\delta\veps_n^{\lambda}/\lambda)$ and $\lim_{\lambda \rightarrow
0}(\delta\varphi_n^{\lambda}/\lambda)$. In practice, however, the
higher orders in Breit interaction are usually small enough to be
neglected even for $\lambda=1$ and there is no need in calculating
these limits.

\paragraph{Correlation effects.}
The usual accuracy of the DF approximation for binding energies of
heavy atoms is of the order of 10\%, while the Breit corrections
are usually about 1\% or less. Therefore, there is no point in
calculating Breit corrections if one is not going to account for
correlations in some way. It is usually convenient to distinguish
between valence-valence and valence-core, or core-core
correlations. The easiest and straightforward way to treat the
former is to use the configuration interaction (CI) method, while
the latter are treated more efficiently within the many-body
perturbation theory (MBPT) in residual Coulomb interaction
\cite{DFK}. Below we discuss how to include Breit interaction in
the CI and the MBPT calculations.

If the CI is done in the frozen core approximation, then as we have
said above, there is no need to include Breit corrections to the
two-electron interaction, but it is necessary to include Breit
corrections to the core potential. It is also important that core
orbitals are found from the DFCB equations.

If the MBPT in residual Coulomb interaction is used to account for
valence-core and core-core correlations, one may need to include
Breit corrections to corresponding diagrams. Generally speaking,
there are two types of corrections: ``direct'' and ``indirect''
ones. The former corresponds to the substitution of the residual
Coulomb interaction with the Breit interaction in the MBPT
expressions, and the latter corresponds to the use of the DFCB
equations as a zero order approximation. The direct corrections are
suppressed because the largest Breit radial integrals correspond to
the virtual excitations from the innermost core shells (see above)
and these excitations correspond to the huge energy denominators.
Therefore, one can neglect them without significant loss of
accuracy. The indirect corrections are much simpler: they are
accounted for simply by the use of the DFCB basis set instead of
the DFC one.

We see that dominant Breit corrections come from the solution of
the DFCB equations for atomic orbitals. After that these orbitals
can be used as a basis set for the CI and the MBPT calculations. On
the CI stage one has to include Breit potential of the core
explicitly when calculating one-electron integrals for valence
electrons, while in the MBPT calculations the direct Breit
corrections can be neglected altogether. That significantly
simplifies calculations and allows to use standard CI and MBPT
codes without significant changes.

Up to now we have focused on the calculations of atomic spectra. If
it is necessary to calculate some other atomic observables, the
general calculation scheme remains the same \cite{DKPF}. However,
the MBPT part should be extended and some new types of the MBPT
corrections may appear. Also, it may be necessary to solve the
random-phase approximation (RPA) equations for the operator of
interest. For example, if we calculate hyperfine structure (HFS)
constants, we have to solve the RPA equations for the operator of
the hyperfine interaction. Again there are direct and indirect
Breit corrections to these equations. The latter are easily
included in the same way as above. The former are important only
for operators which are singular at the origin and, thus, have
large matrix elements for the innermost core electrons. This is the
case for the hyperfine interaction, while electric dipole operator
gives the opposite example.

\paragraph{Numerical results.}
Here we present some results for Cs and Tl. These two atoms are
interesting because they are used in precision measurements of
parity nonconservation and high accuracy atomic calculations are
necessary to compare these measurements with Standard model
predictions. The nuclear charge $Z$ for Tl is much larger than for
Cs, so we can also see how Breit corrections grow with the nuclear
charge.

In \tref{tab1} we present results of the DF calculations of binding
energies for Cs in the Coulomb and Coulomb-Breit approximations. We
also give results of the perturbative calculations of the Breit
corrections for comparison. It is seen that the relaxation of the
core significantly reduces Breit corrections to the valence
energies.

In \tref{tab2} the direct Breit corrections to the HFS constants
are given. As we have mentioned above, here one has to calculate
Breit correction to the RPA vertices in addition to the corrections
to the orbitals. In notations of Derevianko \cite{Der} the latter
corresponds to the one-particle correction and the former
corresponds to the two-particle one. It is seen again that
corrections obtained in the self-consistent approach are much
smaller than the corresponding first order perturbative
corrections. That is caused by the screening of the Breit
interaction by the core electrons which is neglected in the
perturbative approach. Note, that screening effect is particularly
important for the $s$-wave.

The difference between our calculations of the first order Breit
corrections to the HFS constants and the results of the paper
\cite{Der} should be due to the retardation term in \eref{II_3}.
This difference is about 10\% for the $p$-wave and 25\% for the
$s$-wave. That confirms that this term is significantly smaller
than the magnetic one, but it can become important when Breit
correction becomes larger or the accuracy of the calculations is
improved.


On the next stage we calculated correlation corrections following
the method described in \cite{DFK,DKPF}. These corrections include
the self-energy correction to the Hartree-Fock potential and the
structural radiation corrections to the HFS operator. The latter
contribution is small and we did not calculate Breit corrections to
it. Note that the indirect Breit corrections were calculated for
the self-energy and the RPA contributions, while the direct Breit
corrections for the self-energy were neglected. The results of
these calculations are given in \tref{tab3}. It is seen that direct
and indirect Breit corrections to the HFS constants are comparable.
That means that it is important to include Breit corrections when
electronic correlations are considered.

Calculations for Tl are much more complicated and require large CI
for three valence electrons in addition to the MBPT (see
\cite{DFK,DKPF} for details). The Breit corrections appear to be
much more important for Tl than for Cs. Here they are not
negligible already when the energy spectrum is considered (see
\tref{tab4}). In particular, the fine structure splitting of the
ground state is changed by 1\%. Another difference from Cs is that
the screening effect here is relatively weaker, but still
significant. For example, the first order Breit correction to the
fine splitting of the ground state is $-130$~cm$^{-1}$ while the
DFCB value is $-81$~cm$^{-1}$.

Comparison of the results of the CI calculations from \tref{tab4}
with the final results, which include the MBPT corrections,
demonstrates that Breit contributions to the MBPT part constitute
about 10\% of the total Breit corrections to the transition
frequencies. That is consistent with the overall role of the MBPT
corrections in the final answer where they also contribute about
10\%.

\paragraph{Conclusions.}
We see that Breit interaction is important in precision
calculations of heavy atoms. It is sufficient to include only
(exchange) Breit potential of the core and neglect valence-valence
Breit interaction. It is important to allow core relaxation to
avoid overestimation of Breit corrections to energies and orbitals
of valence electrons. It is also important to calculate Breit
corrections to the RPA part and to the MBPT part when one
calculates the HFS constants or other similar observables.

\paragraph{Acknowledgments.}
We are grateful to Dzuba, Labzowsky, Mosyagin, Titov, and
Trzhaskovskaya for helpful discussions of the role of Breit
interaction in heavy atoms. This work was partly supported by RFBR
grants No 98-02-17663 and 98-02-17637.


\begin{table}
\caption{Binding energies for Cs (au) in Dirac-Fock-Coulomb (DFC)
and Dirac-Fock-Coulomb-Breit (DFCB) approximations. The last raw
gives the one-particle first order Breit (FOB) correction in the
perturbative approach. }

\label{tab1}

\begin{indented}
\item[]\begin{tabular}{lcccccc}
\br
&$6s_{1/2}$ &$6p_{1/2}$ &$6p_{3/2}$
&$7s_{1/2}$ &$7p_{1/2}$ &$7p_{3/2}$\\
\mr
DFC  &0.127368 &0.085616 &0.083785 &0.055187 &0.042021 &0.041368 \\
DFCB &0.127358 &0.085577 &0.083768 &0.055183 &0.042008 &0.041362 \\
FOB  &0.127217 &0.085537 &0.083726 &0.055146 &0.041993 &0.041347 \\
\br
\end{tabular}
\end{indented}
\end{table}
\begin{table}
\caption{The direct Breit corrections to the hyperfine structure
constants in Cs(MHz). Corrections to the orbitals are calculated
perturbatively (FOB) and self-consistently (DFCB). Direct
corrections to the RPA are calculated as discussed in the text. }

\label{tab2}

\begin{indented}
\item[]\begin{tabular}{lrrrrrr}
\br
& \multicolumn{1}{c}{DFC}
& \multicolumn{2}{c}{FOB}
& \multicolumn{1}{c}{DFCB}
& \multicolumn{2}{c}{RPA} \\
&&\multicolumn{1}{c}{this work}
&\multicolumn{1}{c}{\cite{Der}}
&&\multicolumn{1}{c}{this work}
&\multicolumn{1}{c}{\cite{Der}}\\
\mr
$A_{6s_{1/2}}$ & 1423.8 &$-10.7$ &$-8.14$ &$-1.2$ & 3.1 & 3.50\\
$A_{6p_{1/2}}$ &  160.9 &$-1.9$  &$-1.58$ &$-1.1$ & 0.7 & 0.73\\
$A_{6p_{3/2}}$ &   23.9 &$-0.2$  &   ---  &$-0.1$ & 0.1 & --- \\
$A_{7s_{1/2}}$ &  391.2 &$-2.4$  &$-1.80$ &$ 0.0$ & 0.9 & 0.96\\
$A_{7p_{1/2}}$ &   57.6 &$-0.6$  &$-0.54$ &$-0.4$ & 0.2 & 0.26\\
$A_{7p_{3/2}}$ &    8.6 &$ 0.0$  &   ---  &$ 0.0$ & 0.0 & --- \\
\br
\end{tabular}
\end{indented}
\end{table}
\begin{table}
\caption{Final results with the MBPT corrections for the HFS
constants for Cs (MHz) in Coulomb and Coulomb-Breit approximations.
The definitions of the direct Breit (DB) and the indirect Breit
(IB) corrections are given in the text. Experimental result are
taken from \cite{AIV,GWW}. }

\label{tab3}

\begin{indented}
\item[]\begin{tabular}{lcrrrr}
\br
& \multicolumn{1}{c}{Coulomb}
& \multicolumn{1}{c}{DB}
& \multicolumn{1}{c}{IB}
& \multicolumn{1}{c}{Total}
& \multicolumn{1}{c}{Exper.} \\
\mr
$A_{6s_{1/2}}$ & 2312 & $ 1.9$ & $ 0.9$ & 2315 & 2298 \\
$A_{6p_{1/2}}$ & 296  & $-0.4$ & $-0.4$ & 295  & 292  \\
$A_{6p_{3/2}}$ & 55.3 & $ 0.0$ & $-0.1$ & 55.2 & 50.3 \\
$A_{7s_{1/2}}$ & 549  & $ 0.9$ & $-0.9$ & 549  & 546  \\
$A_{7p_{1/2}}$ & 94.0 & $-0.2$ & $-0.1$ & 93.6 & 94.3 \\
$A_{7p_{3/2}}$ & 18.2 & $ 0.0$ & $ 0.0$ & 18.2 & 18.2 \\
\br
\end{tabular}
\end{indented}
\end{table}
\begin{table}
\caption{Calculation of the spectrum of Tl in the Coulomb (C) and
the Coulomb-Breit (CB) approximations (in cm$^{-1}$). The DF
equations were solved in the $V^{N-1}$ approximation. The CI was
done for three valence electrons and final calculation included
core-valence and core-core correlations within the CI+MBPT method.
}

\label{tab4}

\begin{indented}
\item[]\begin{tabular}{lrrrrrrr}
\br
& \multicolumn{2}{c}{DF}
& \multicolumn{2}{c}{CI}
& \multicolumn{2}{c}{Final}
& \multicolumn{1}{c}{Exper.} \\

& \multicolumn{1}{c}{C}
& \multicolumn{1}{c}{CB}
& \multicolumn{1}{c}{C}
& \multicolumn{1}{c}{CB}
& \multicolumn{1}{c}{C}
& \multicolumn{1}{c}{CB}
& \multicolumn{1}{c}{\cite{More}} \\
\mr
$6p_{1/2}$ &    0 &    0 &    0 &    0 &    0 &    0 &    0 \\
$6p_{3/2}$ & 7186 & 7105 & 7066 & 6988 & 7878 & 7780 & 7793 \\
$7s_{1/2}$ &22713 &22597 &24767 &24660 &26596 &26474 &26478 \\
$7p_{1/2}$ &29546 &29434 &32210 &32108 &34109 &33993 &34160 \\
$7p_{3/2}$ &30465 &30346 &33159 &33048 &35119 &34994 &35161 \\
$6d_{3/2}$ &      &      &33970 &33851 &36175 &36041 &36118 \\
$6d_{5/2}$ &      &      &34029 &33909 &36250 &36115 &36200 \\
\br
\end{tabular}
\end{indented}
\end{table}
\end{document}